\documentclass{aa}  

\usepackage{graphicx}
\usepackage[usenames,dvipsnames,svgnames]{xcolor}
\usepackage{url}             
\usepackage{txfonts}
\usepackage{makecell}
\usepackage{lineno}
\graphicspath{
    {./}
}
\newcommand{\BE}{\begin{equation}}
\newcommand{\EE}{\end{equation}}
\newcommand{\BA}{\begin{eqnarray}}
\newcommand{\EA}{\end{eqnarray}}
 \newcommand{\fig}[1]{Fig.~\ref{fig_#1}}
 
 \newcommand{\figs}[2]{Figs.~\ref{fig_#1} and \ref{fig_#2}}
 
 \newcommand{\sect}[1]{Sect.~\ref{sect_#1}}
 \newcommand{\sects}[2]{Sects.~\ref{sect_#1} and~\ref{sect_#2}}
 \newcommand{\eq}[1]{Eq.~(\ref{eq_#1})}

 \newcommand{\tab}[1]{Table~\ref{tab_#1}}

\renewcommand{\vec}[1]{{\bf #1}}

\newcommand{\eg}{\textit{{\it e.g.},}}

\newcommand{\ie}{\textit{i.e.}}
\newcommand{\insitu}{in situ}

\newcommand{\img}[2]{\includegraphics[scale=#1]{{#2}.png}}
\def \Vmc       {$V_{mc}$}              
\def \sigsh     {$\sigma_{sh}$}         
\def \nGCR      {$n_{GCR}$}             
\def \np        {$n_{p}$}               
\def \tauFD     {$\tau_{FD}$}           
\def \Tp        {$T_{p}$}               
\def \Vsh       {$V_{sh}$}              
\def \Bsh       {$B_{sh}$}              
\def \dtsh      {$\Delta t_{sh}$}       
\def \Bmc       {$B_{mc}$}              
\def \dtmc      {$\Delta t_{mc}$}       

\begin{document} 
    \title{Superposed epoch study of ICME sub-structures \\ 
    near Earth and their effects on galactic cosmic rays}

    \titlerunning{Superposed epoch study of ICMEs and their effects on galactic cosmic rays} 
    \authorrunning{J. J. Mas\'\i as-Meza et al.}

    \author{
        Mas\'\i as-Meza, J.J.\inst{1}
        \and Dasso, S. \inst{2,3}
        \and D\'emoulin, P. \inst{4}
        \and Rodriguez, L. \inst{5}
        \and Janvier, M. \inst{6}
    }

    \institute{
        Departamento de F\'\i sica - IFIBA, Facultad de Ciencias Exactas y Naturales, Universidad de Buenos Aires, 1428 Buenos Aires, Argentina. \email{masiasmj@df.uba.ar}
        \and
        Instituto de Astronom\'\i a y F\'\i sica del Espacio, UBA-CONICET, CC. 67, Suc. 28, 1428 Buenos Aires, Argentina. 
        \and
        Departamento de Ciencias de la Atm\'osfera y los Oc\'eanos and Departamento de F\'\i sica, Facultad de Ciencias Exactas y Naturales, Universidad de Buenos Aires, 1428 Buenos Aires, Argentina. 
        \and
        Observatoire de Paris, LESIA, UMR 8109 (CNRS), F-92195 Meudon Principal Cedex, France 
        \and
        Solar–Terrestrial Center of Excellence – SIDC, Royal Observatory of Belgium, Av. Circulaire 3, 1180, Brussels, Belgium
        \and
        Institut d'Astrophysique Spatiale, UMR8617, Univ. Paris-Sud-CNRS, Universit\'e Paris-Saclay, B\^atiment 121, 91405 Orsay Cedex, France
     }

   \date{Received January, 2016; accepted May 12, 2016}

  \abstract
   {
Interplanetary coronal mass ejections (ICMEs) are the interplanetary manifestations of solar eruptions. 
The overtaken solar wind forms a sheath of compressed plasma at the front of ICMEs.
Magnetic clouds (MCs) are a subset of ICMEs with specific properties (\eg\ the presence of a flux rope).
When ICMEs pass near Earth, ground observations indicate that the flux of galactic cosmic rays (GCRs) decreases. 
}
   {
The main aims of this paper are to find:
common plasma and magnetic properties of different ICME sub-structures, 
and which ICME properties affect the flux of GCRs near Earth.
   }
   {        
We use a superposed epoch method applied to a large set of ICMEs observed \insitu\ by the spacecraft ACE, between 1998 and 2006. 
We also apply a superposed epoch analysis on GCRs time series observed with the McMurdo neutron monitors.
}
    {We find that slow MCs at 1~AU have on average more massive sheaths. 
    We conclude that it is because they are more effectively slowed down by drag during their travel from the Sun. 
    Slow MCs also have a more symmetric magnetic field and sheaths expanding similarly as their following MC, while in contrast, fast MCs have an asymmetric magnetic profile and a 
sheath in compression.
   In all types of MCs, we find that the proton density and the temperature, as well as the magnetic fluctuations can diffuse within the front of the MC due to 3D reconnection. 
   Finally, we derive a quantitative model which describes the decrease of cosmic rays as a function of the amount of magnetic fluctuations and field strength.
}
   {The obtained typical profiles of sheath/MC/GCR properties corresponding to slow, mid, and fast ICMEs, can be used for forecasting/modelling these events, 
   and to better understand the transport of energetic particles in ICMEs. They are also useful for improving future operative space weather activities.}


   \maketitle

%

\section{Introduction}
\label{sect_intro}

Coronal mass ejections (CMEs) are associated with magnetic instabilities occurring in the solar corona, and they are 
expelled during solar eruptive flare events.
As a consequence of these instabilities, large quantities of plasma and magnetic fields are expelled into the interplanetary space (IP), which can be observed by coronagraphs as CMEs. 
When their interplanetary manifestations (interplanetary CMEs or ICMEs) arrive at Earth, the observed intensity of energetic particles (\eg\ Galactic Cosmic Rays, GCR) is modified.
When they arrive at other planets (\eg\ Mars) energetic particle fluxes can be also modified.
ICMEs 
can also produce perturbations in the magnetosphere, triggering geomagnetic storms.
A subset of ICMEs include a Magnetic Clouds (MC) which is characterised by a low proton temperature and an enhanced magnetic field intensity $|\vec{B}|$ with a smooth rotation of its vector components, resembling that of a flux rope \citep{burlaga81}. 
The solar wind following an MC is expected to be perturbed, with characteristics similar to a wake.

Typically, MCs travel faster than the local Alfv\'en waves in the solar wind reference frame, producing a fast MHD shock ahead of them.
This shock produces an intermediate region of compressed plasma between the shock interface and the MC leading edge. 
This region is characterised by high temperatures produced by the conversion of macroscopic to thermal energy at the shock, and therefore it generally presents high plasma $\beta$ values.
Typically, sheath regions also present large magnetic intensity and a high level of magnetic fluctuations.

Fluctuations around the shocks in the solar wind are 
observed after the shock (downstream) as typically observed in fluids, but also upstream (before the arrival of the shock).  
This last case is mainly due to beam instabilities that are induced by shock-accelerated particles. 
These instabilities can generate waves in the upstream region
\citep[\eg\ ][]{blancocano_etal11,wang15,strumik15}, so that an increased level of fluctuations is expected before and after the shocks associated with ICMEs.

During their propagation away from the Sun, MCs interact with the plasma and magnetic field encountered in the interplanetary medium.
This interaction can induce reconnection between the sheath and the MC magnetic fields.  
This implies changes of the magnetic connectivity of the flux rope \citep{mccomas88}, and a consequent peel off (erosion) of magnetic flux from the leading edge of the MC.
This also implies the formation of a 'back region' which involves field lines that were part of the MC before the erosion \citep{dasso06,Ruffenach15}. 
The back region is located in the MC wake and it typically has mixed plasma and magnetic field properties of ambient solar wind and MC. 

Sheaths in front of MCs significantly differ from planetary magneto-sheaths mainly because ICMEs not only propagate but also expand into the IP medium \citep[\eg\ ][]{Demoulin09,Gulisano10}.
In particular, the lateral deflection of the solar wind away from the nose of an MC is reduced due to the expansion, and the solar wind tends to pile up in front of an MC instead of flowing around it \citep{Siscoe08}.
This effect is more important near the corona, where the expansion is stronger, than in the interplanetary medium \citep{Das11}.
Moreover, the drainage toward the sides can be enhanced due to magnetic reconnection between the magnetic field of the sheath and the flux rope.

Summarising, a typical ICME is expected to be formed by the following sub-structures: 
slightly enhanced level of fluctuations (upstream waves turbulence), shock (driven by the MC), sheath (shocked, compressed, heated and turbulent material),  
MC (flux rope), back (mixed flux rope and solar wind plasma when erosion has occurred), and wake (perturbed solar wind after the back region of the flux rope).

Transient structures in the heliosphere affect the transport of energetic particles in the solar wind \citep[\eg\ ][and references therein]{masson12}. 
On the one hand, acceleration at shocks driven by ICMEs is the main mechanism involved in the production of gradual energetic particle events in the inner heliosphere \citep[\eg\ ][]{vainio09}. 
On the other hand, the decrease of the flux of energetic particles over a huge range of energies is associated with shocks and ICMEs. 
At lower energy, this is observed \insitu.  
For higher energies (\eg\ galactic cosmic rays, GCRs) this is observed at ground level by neutron monitors \citep[\eg\ ][]{Simpson54}, by muon telescopes \citep[\eg\ ][]{arunbabu15}, or by water-Cherenkov detectors \citep[]{auger11_scalers,Dasso12,Asorey16}.

In particular, the passage of ICMEs and their associated shocks have important effects on GCRs. 
Indeed, a Forbush decrease (FD) of galactic particles is typically observed during several days and in association with the passage of an ICME \citep[\eg\ see the review,][]{cane00}. 
Additionally, a smaller amplitude and a higher-frequency variability in the GCR intensity compared with classical FDs can also be observed \insitu\ near Earth \citep[\eg\ ][]{mulligan09}. 
Furthermore, a decrease in the abundance of energetic particles associated with FD has also been \insitu\ observed by the Mars Science Laboratory's rover Curiosity, during its cruise phase from Earth to Mars \citep{guo_etal15}.  

The full structure of ICMEs can perturb the transport of GCRs and has important effects both locally and globally on the density of GCRs.
These effects are associated respectively with 
  (1) strong changes of the local properties of the solar wind turbulence (mainly in the sheath) 
  which consequently affect the diffusion coefficients of these energetic particles and 
  (2) the presence of structures with smooth closed \vec{B} field lines with typically high $|\vec{B}|$ values inside MCs, which hardly allow diffusion transport across \vec{B} \citep[\eg\ ][]{krittinatham09}.

From a statistical study of interplanetary properties of ICMEs and muon data for cutoff rigidities 
between 14 and 24 GV (using the GRAPES-3 muon telescope), \cite{arunbabu15} found that the enhancement of the interplanetary magnetic field inside the sheath regions is strongly associated with the observed FD profile.
They concluded that FDs are mainly caused by the cumulative diffusion of protons across $\vec{B}$ in the sheath. 
However, it remains quantitatively unclear how FDs can be modelled, in particular how one can quantify the importance of $|\vec{B}|$ and the level of turbulent fluctuations. 

In this paper we characterise the mean properties of MCs at 1 AU and their relation to GCR transport.
We apply the superposed epoch analysis method on \insitu\ solar wind data and ground-based neutron monitor observations.
This method is powerful because it emphasises the common properties and removes peculiarities of some events. 
It is similar to the superposed epoch developed by \cite{lepping03b}, where they studied 19 MCs from 1995 to 1998 and created a profile of  magnetic field and plasma parameters of MCs, and even more to the method developed by \cite{Yermolaev15} where they studied and compared the profiles of ejecta, MCs, their sheaths and corotating interaction regions.
Our method is also similar to the one recently developed in \cite{rodriguez16}, to analyse plasma, magnetic and composition properties, focussing on the physical mechanisms for the formation of structures inside and at the rear of MCs.
It is also close to the method used by \cite{badruddin_etal16}, where the response of GCRs to corotating interaction regions and ICMEs was studied with a focus on their interfaces with the ambient solar wind.

Finally, our study is related to the one of \cite{belov_etal15} since they model the response of GCRs inside MCs as a function of the interplanetary conditions (plasma speed and magnetic field), and geomagnetic properties (Dst index).
In particular, they report the presence of 
a local minimum of GCR inside MCs for strong-field events.

In \sect{data_and_events} we present the ICMEs studied and the data used to analyse them.
In \sect{superposed_epoch} the superposed epoch method, applied to our sample, provides the typical profiles of magnetic field and plasma parameters in ICMEs containing an MC.
In \sect{splits_with_V} we further investigate how these profiles depend on the strength of the MC, in particular its mean velocity. 
For that we split our sample in slow, mid and fast events.
The association of the event properties with the observed flux of GCRs is presented in \sect{forbush}. 
These results allow us to propose a novel quantitative model to describe the temporal evolution of GCR flux using parameters observed in the solar wind.
Finally, in \sect{conclusions}, we present our conclusions.

\section{Data and events selection}
\label{sect_data_and_events}

\begin{figure*}[ht]
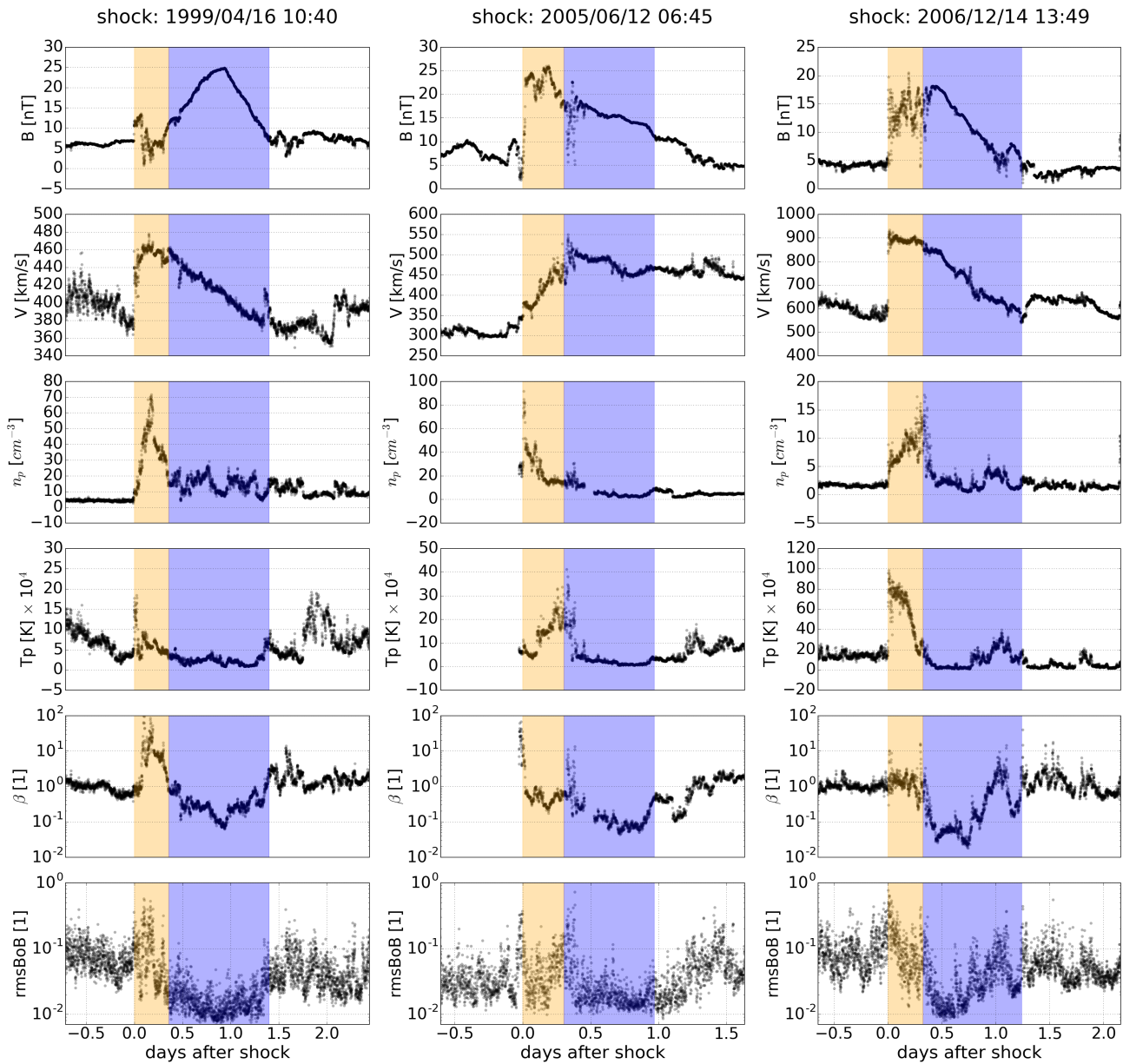

    \sidecaption    
    \img{0.25}{example_01}   
    \img{0.25}{example_03}   
    \img{0.25}{example_02}   
    \caption{
    Time profiles for different observables, from top to bottom: magnetic field $B$, bulk velocity $V$, proton density \np, proton temperature \Tp, plasma beta $\beta$, normalised magnetic-fluctuation density $rmsBoB$ (\eq{rmsBoB}), and absolute magnetic fluctuation $rmsB$ (\eq{rmsB}).
    We show three different events; one per column.
Each of them belongs to each of the groups analysed in \sect{splits_with_V}; from left to right, the events belong to the slow, intermediate, and fast groups.  
    The date of the shock arrival is shown at the top of each column in UT.
    The passage of the sheath is marked in orange, and the passage of the MC in blue.
    The fastest event shows stronger compression at the MC front, with respect to the others.
    The slowest event presents higher density values within the sheath. See \sect{quantities_examples} for a description of the main important features shown in each panel.
    }
\label{fig_examples}
\end{figure*}

In this section we present the data used for the analysis, the sample of studied events, and the observed physical quantities we explore.

\subsection{Events selection}
\label{sect_event_selection}

We use data of interplanetary magnetic field and plasma from the MAG and SWEPAM experiments onboard the ACE spacecraft \citep{ACE98, mccomas98_ace}, using a 64 seconds time cadence, which are available at \url{http://www.srl.caltech.edu/ACE/ASC/level2/}.
The intensity of GCRs is analysed using data from McMurdo neutron monitors 
with one hour time cadence; these data are available at \url{http://neutronm.bartol.udel.edu/~pyle/bri_table.html}.

We studied events taken from the list of \cite{Richardson10}, including only ICMEs flagged as MCs from 1998 to 2006, \url{http://www.srl.caltech.edu/ACE/ASC/DATA/level3/icmetable2.htm}.
Specifically, we use only those events satisfying the following conditions: 
a) ICMEs with flag \texttt{2} in the list, which means that they are included in the WIND MC list compiled by R.P. Lepping, \url{http://lepmfi.gsfc.nasa.gov/mfi/mag_cloud_S1.html}.
b) no multiple MCs, to avoid complex structures involving interactions of different events.
c) MCs with a sheath and with an associated shock.

We selected the shock events from the shock list of \cite{wang10}.
We intersect both ICME and shock catalogs, only selecting ICMEs for which associated shocks are less than 3 hours before the ICME starting time.
The above procedure defines 44 events, and we summarise their main characteristics in \tab{selec_events} (on line version).


Because the times listed in the table of \cite{Richardson10} are referenced at Earth and our data are referenced at L1, we shift these listed times at L1, where the ACE data are observed.  
We use a global time shift as defined by the time difference of the shock observed at L1 and at Earth.
We made this shift for each ICME by identifying the shock discontinuity in the time series of the magnetic field and plasma parameters.
The time shifts are written in the column $5$, shown in \tab{selec_events}.

\subsection{Analysed physical quantities}
\label{sect_quantities_examples}

In this section we describe some characteristics of three individual events as a context for further discussions of our results on average profiles.
In \fig{examples} we show three MCs: one with a slow bulk velocity, one with a mid velocity value, and a fast one (shown on the left, central, and right column, respectively).

The upper panels show that the magnetic field strength has a comparable magnitude for the three selected MCs ($B_{\rm max} \approx 20$ nT). 
The compression of the magnetic field (enhanced intensity) is present in the MC sheath, especially for the mid and fast MCs. 
The faster MC presents also the most asymmetric B profile while the slow MC present a very symmetric B profile. 

The second row of panels shows a gradient in the velocity profile.  
This typical velocity profile is associated with the MC expansion as it moves away from the Sun.
The slow and fast events present a very different change on the velocity during the MC passage (so that $\Delta V =V_{end}-V_{start}$ is larger for the fast event compared to the slow one, a result consistent with the results of \cite{Gulisano10} for MCs observed in the inner heliosphere).  
Finally, we note that the event chosen with intermediate speed looks perturbed with a break in the velocity profile around the half time of the MC.

In the third and fourth rows, we see enhanced proton density \np\ and temperature \Tp\ inside the sheaths, as expected because these regions are formed of compressed and heated plasma behind the shock.

In the fifth row, we show the plasma $\beta$. 
Its value is estimated with the same method as in the OMNI database, \ie\ by assuming that the temperature of alpha particles, $T_\alpha$, is proportional to the proton temperature with: $T_\alpha = 4 T_p$, and that the electron temperature is constant $T_e = 1.4 \times 10^5 K$ (see \url{http://omniweb.gsfc.nasa.gov/ftpbrowser/magnetopause/Reference.html}).
These examples show that $\beta$ can be greater than 10 inside the sheath, while inside the MC, $\beta \ll 1$ as expected \citep[\eg\ ][]{Dasso05}.

We also present the fluctuations of the unit vector of the magnetic field $\vec{B}$, which we call the normalised magnetic fluctuations 
(i.e., rms of B over its magnitude B) and write it as $rmsBoB$ hereon.
We define it with:
   \begin{align}
    \centering
    rmsBoB(t) &= rmsB(t) / B(t) \label{eq_rmsBoB} \\
    rmsB(t)   &= \sqrt{\sum_{i=1}^{3} <(B_i- \langle B_i \rangle)^2>} 
    \label{eq_rmsB}
   \end{align}
where $rmsB$ is computed in time windows of 16 seconds, using a high time cadence of 3 vectors per second, and the mean value of each component $\langle B_i \rangle$ is computed inside this time window (as provided by the ACE webpage).
$rmsBoB$ is much lower inside MCs, than outside (sixth row of \fig{examples}).
In the next section, we will see that this magnitude is a very robust quantity within MCs.

%
%

The magnetic fluctuations $rmsB$ are shown in the seventh (bottom) row. 
They have maximum values within the sheath, while inside MCs it is roughly similar to the values observed before the shock arrival.

Besides these expected features in these three cases, we obtain below typical profiles of ICME events, and quantify them.
This is the motivation to implement a superposed epoch analysis in the next section.

\begin{figure*}
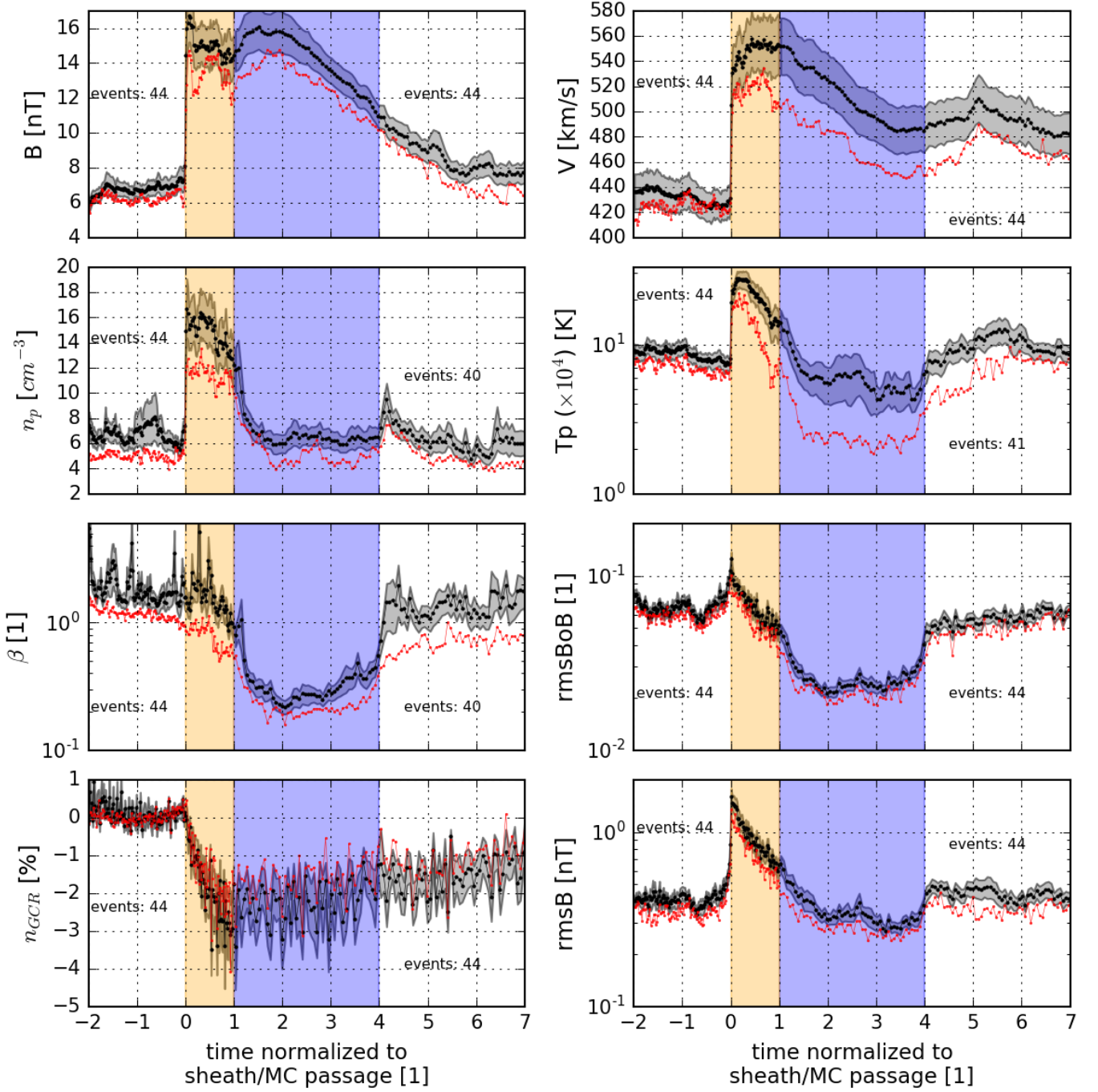

\centering
    \img{0.9}{figs_all.global}
    \caption{
    Superposed epoch for different quantities derived with the method described in \sect{superposed_method}.   
    The sheath is shown with an orange background colour (time range $0<t<1$) and the MC with a blue background colour ($1<t<4$).
    The time $t$ is normalised with the sheath duration for $t<1$ and with the MC duration for $t>1$ (the MC duration is three times longer than the sheath duration as observed in average).
    The black dots, connected with a line, are averaged values for each bin on the time axis, the grey band represents the error of the mean and the red line represents the median values in each time bin.
    The average profiles were computed filtering those events that had less than 20\% of data-gaps in the transient structure (MC or sheath).
    The number of events after filtering is shown in each panel.
The panels represent: magnetic field $B$, bulk velocity $V$, proton density \np, proton temperature \Tp, plasma beta $\beta$, normalised magnetic-fluctuation density $rmsBoB$ (\eq{rmsBoB}), GCR intensity normalised to the pre-ICME level \nGCR\ and absolute magnetic fluctuation $rmsB$, \eq{rmsB}.
    A clear discontinuity of the magnetic fluctuation intensity, $rmsB$, is present at the shock position (front of the sheath).
    The normalised magnetic fluctuations, $rmsBoB$, inside the MC are significantly lower than in the ambient solar wind (by a factor of $\simeq$ 3).
}
    \label{fig_avrg_profiles}
\end{figure*}

\section{Superposed epoch for the sheath and MC structures}
\label{sect_superposed_epoch}

\subsection{Method}
\label{sect_superposed_method}

The main aim of the superposed epoch is to obtain an average profile by taking a sample of individual profiles.
In order to obtain this average, each individual profile must have the same number of data points or bins in the time dimension.
However, the primitive data have different durations for different events so we implement a re-binning such that within each MC we end up with 50 time bins. 
Then, the data within each bin are averaged to a single value per bin. 
In the case when a primitive time series has a data gap in more than 20$\%$ of the MC structure, we discard it from the average profile.
We also re-bin the data obtained after the MC, over the same time interval and with the same number of bins.
Next, we average the data associated to the different events, by averaging bin by bin each quantity ($B$, $V$, $\beta$, etc), which finally builds the average profiles. 
   
We repeat this procedure to obtain average profiles of the sheath structure.
Again, we resample all the time series to 50 bins inside each sheath.  
Since the number of events having more than 20$\%$ of gaps are not necessarily the same for the sheath analysis as for the MC analysis, we have indicated in each graph the total number of cases taken for both the sheath and the MC.
The same time normalisation as for the sheath is used to resample the data before the shock, over twice the time interval of each sheath. 

We next build a combined profile which shows an average profile for: the background solar wind, the sheath, the MC and the MC wake. 
We select the time origin at the shock.  
Finally, we set the temporal lengths of the MC to sheath to a ratio 3:1 to better represent the typical relative durations of each structure. 

In \fig{avrg_profiles}, we show the average profiles (black lines) with the associated errors of the means (grey bands), and the median values (red lines).
The difference between the median and the mean profiles is a proxy of how non-symmetric the distributions of the quantities are.
Indeed, in the central part of MCs, the observables are distributed in a log-normal manner \citep{rodriguez16}.
Similar log-normal distributions also have been found for ICME observables in \cite{guo_etal10} and \cite{Mitsakou14}.

\begin{table*}
\centering
\caption{
    Mean values of all quantities (column (1)) for all events (columns (2) and (3)), 
and for each group of events associated with MCs with low, mid and high velocities (columns (4) to (9)).
    These mean values are calculated within sheaths and MCs.
}
\begin{tabular}{c|cccccccccccccccrr}
\hline
& & \multicolumn{2}{c}{All} & &\multicolumn{2}{c}{$V_{mc}^{low}$} & & \multicolumn{2}{c}{$V_{mc}^{mid}$} & & \multicolumn{2}{c}{$V_{mc}^{high}$} \\
\cline{3-4}\cline{6-7}\cline{9-10}\cline{12-13}
quantity                        &&sheath & MC    &&sheath & MC    &&sheath & MC    && sheath & MC    & \\
 (1) && (2) & (3) && (4) & (5) && (6) & (7) && (8) & (9)  \\

\hline
$V$ [km/s]                      && 561   & 525   && 441   & 402   && 520   & 493   &&  723   & 681 & \\
$B$  [nT]                       && 15.4  & 14.3  && 11.4  & 14.0  && 13.4  & 11.9  &&  21.5  & 16.9  & \\
$n_p$ [1$/cm^3$]                && 14.3  & 6.5   && 18.5  & 9.5   && 14.1  & 5.3   &&  10.1  & 4.7   & \\
\Tp\ [K] ($\times 10^4$)        && 22.8  & 6.5   && 8.3   & 2.6   && 14    & 4.3   &&  46    & 12.4  & \\
$\beta$                         && 1.6   & 0.4   && 2.8   & 0.42  && 1.4   & 0.37  &&  0.7   & 0.25  & \\
$rmsBoB$     ($\times 10^{-2}$) && 6.8   & 2.7   && 6.3   & 2.3   && 6.3   & 3.0   &&  7.9   & 2.8   & \\
$rmsB$ [nT]                     && 0.97  & 0.37  && 0.62  & 0.30  && 0.78  & 0.34  &&  1.53  & 0.47  & \\
\hline
\end{tabular}
\label{tab_avr_values}
\end{table*}

\subsection{Results for the average profiles}
\label{sect_profiles}

From the mean profiles in \fig{avrg_profiles}, we can clearly see sharp jumps at the arrival of the fast shock. 
$B$, $V$, $N_p$, and $T$ jump by factors of 
$\sim$100$\%$, $\sim$30 $\%$, $\sim$150$\%$, and $\sim$300$\%$, respectively.
Then, all quantities behave differently whether in sheath or in the MC.
We describe shortly below the main results for each of the panels of \fig{avrg_profiles} (except the bottom left one, which is described in \sect{model_nGCR}).
Some results in this subsection closely agree with those found in the sample of MCs studied in \cite{Yermolaev15,rodriguez16}.

The upper left panel of \fig{avrg_profiles} shows the piled-up magnetic field $B$ in the sheath. $B$ is enhanced by more than a factor 2 with respect to the pre-shock value.
Next, the $B$ profile inside the MC is strongly asymmetric, with its maximum strongly shifted to the front of the MC.
Indeed, due to the flux rope expansion a stronger $B$ is expected at the front compared to the rear since the spacecraft crosses the front earlier during the propagation of the MC than the rear \citep[\eg\ see Figures 3 and 4, and associated explanation in the main text of][]{Demoulin08}.  
This is the so-called aging effect.
However, it was previously shown that expansion is typically not sufficient to explain such a $B$ asymmetry \citep[see the upper panel of Figure 6 and associated main text of][]{Demoulin08}.

The speed profile (upper right panel of \fig{avrg_profiles}) shows that across the shock there is a jump from 420-440 km/s to 520-540 km/s.
The first part of the sheath shows a fast compressing profile, while it later stabilises to be a roughly constant speed up to the boundary with the flux rope.
Inside the MC, the speed is a decreasing time function, in agreement with the expected typical expansion. 
The linear decrease ends before the MC rear boundary. 
This characteristic is present both in individual profiles and in other superposed epoch studies \citep[\eg\ ][]{lepping03b,rodriguez16}.
A peak of $V$ is also present after the MC.

\begin{figure}
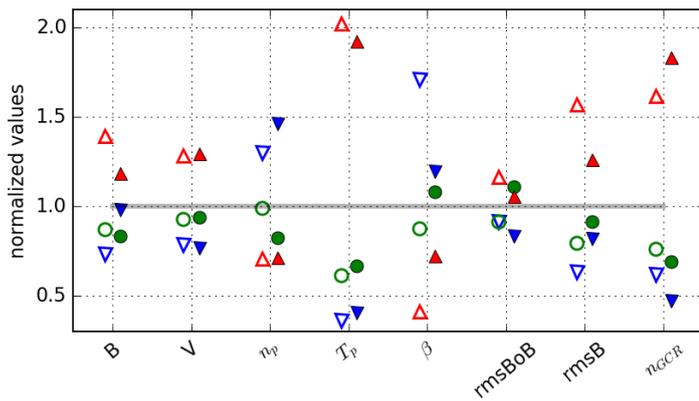
 
    \centering
    \img{0.6}{__paper__deviations_mix_Vmc.group}
    \caption{
    Normalised values of average parameters corresponding to each subgroup (split according to the ordering parameter \Vmc): 
    low \Vmc\ (blue triangle down), mid \Vmc\ (green circle), and high \Vmc\ (red triangle up).
    Empty symbols are for sheaths and filled symbols for MCs.
    For each quantity, we use the mean values of all events for the normalisation of each parameter (grey horizontal line).
    From left to right the parameters are the means of: magnetic field $B$, bulk velocity $V$, proton density \np, proton temperature \Tp, plasma beta $\beta$, normalised magnetic-fluctuation density $rmsBoB$ (\eq{rmsBoB}), absolute magnetic fluctuation $rmsB$ (\eq{rmsB}) and GCR intensity normalised to the pre-ICME level \nGCR.
    }
    \label{fig_Vmc_ordering}
\end{figure}

In the second left panel of \fig{avrg_profiles} the high proton density, \np, in the sheath contrasts with its lower value in the sheath surroundings.
However, \np\ has no discontinuity at the interface between the sheath and the MC.
On the contrary, \np\ starts to decrease from the sheath-MC interface to inside the MC, up to around 15-20\% of the MC size where its value becomes constant.  
This is not an effect of smearing the MC boundary in the averaging since this effect is also present in some individual MCs (\eg\ \fig{examples}).
Finally, a clear density peak is present just after the rear boundary. 
Its physical origin was investigated in \cite{rodriguez16}, where it was shown that several possible mechanisms, such as compression from a fast overtaking stream, or intrinsic mechanisms, such as the eruption conditions at the Sun, could account for this density peak. 

While in the sheath, the magnetic field, velocity and density are roughly constant (besides fluctuations), the proton temperature \Tp\ (right second panel of \fig{avrg_profiles}) has a steep maximum near the shock.  
This is a consequence of the macroscopic to thermal energy conversion produced at the shock. 
Later on \Tp\ progressively decreases towards the MC.
As observed for \np\, the temperature profile is continuous around the sheath-MC interface without a clear break, up to a $\sim$ 15-20 \% of the mean MC size.
Later on, \Tp\ stabilises to a roughly constant value up to almost the end of the MC.

In the density and temperature profiles (\fig{avrg_profiles}), the transition between the sheath region and the 
MC is not as abrupt as expected, considering that the MC environment is much different from the sheath region. 
This transition can be explained as follows.
\cite{gosling_etal95} discuss a mechanism where part of the field lines at the MC periphery are magnetically connected to the outer heliosphere after reconnection occurred.
Case studies \citep{dasso06,dasso07,Feng13} and a statistical analysis of a large sample of events \citep{Ruffenach15} indicate that an MC can suffer magnetic erosion at its front during its travel in the heliosphere, forming a back region with mixed properties of both MC and ambient solar wind \citep{dasso06,Ruffenach12}.
Thus, because MCs are 3D structures, it is possible that this erosion only happens at various specific locations along the flux rope.
Then, the transition region between the sheath and the MC, explored by the spacecraft, can be magnetically connected to another region where erosion has already occurred. 
This gives a possibility for solar wind material and heat to diffuse along the reconnected field lines, which leads to a smoother profile at the interface between those sub-structures.  
In this paper we refer to this mechanism as '3D reconnection'.

\def \sclfg {0.22}
\begin{figure*}[ht]
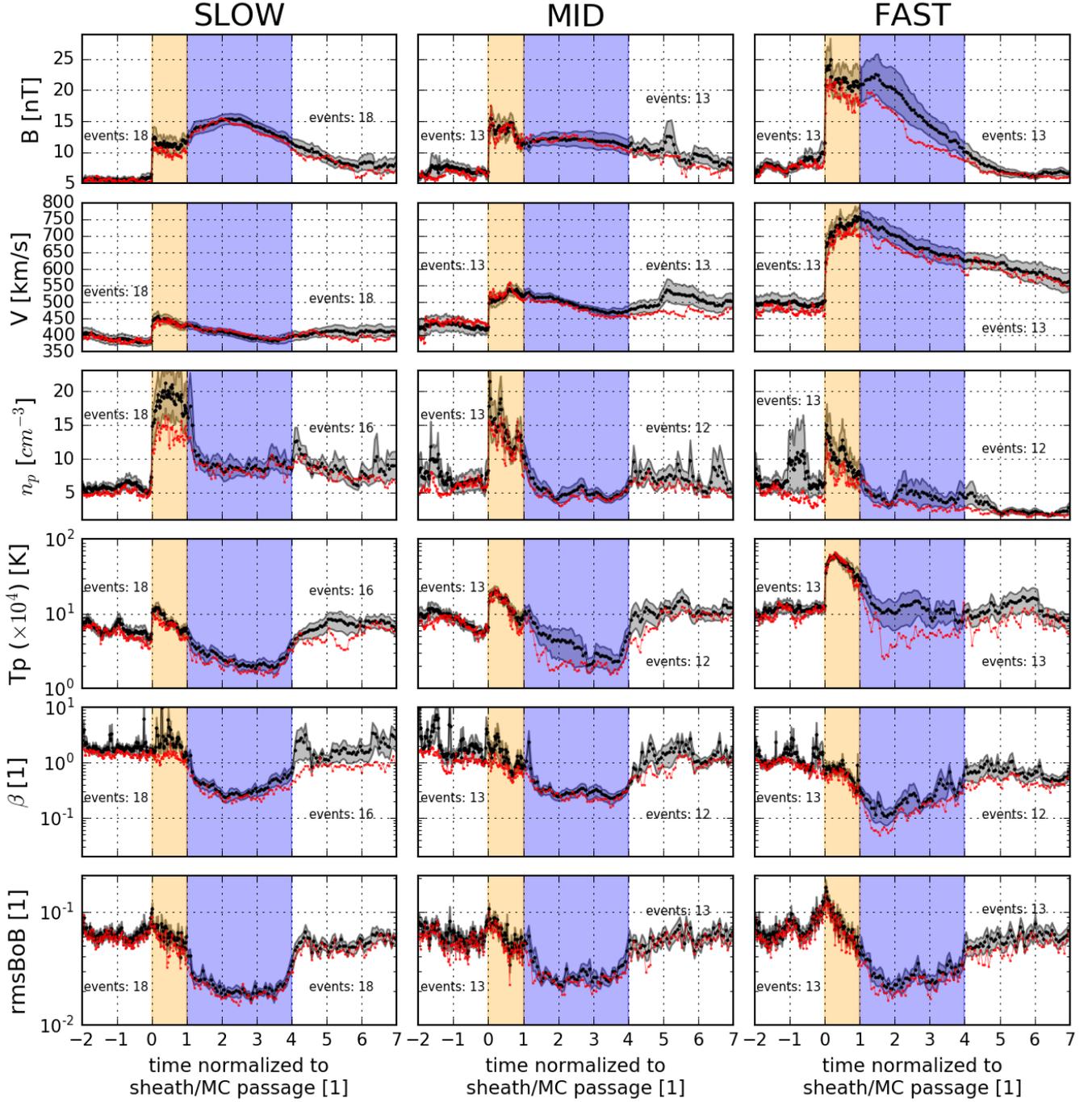

    \sidecaption
    \img{0.8}{figs_splitted_1}
    \caption{
    Average profiles (black lines) distributed in three columns for the three subsets of events. Left: slow ($V<450$ km/s), middle: medium ($450$ km/s$<V<550$ km/s), and right: fast MCs ($V>550$ km/s).
    The black line represents the average value, the grey band represents the error of the mean and the red line represents the median values in each time bin.
    As in \fig{avrg_profiles}, for $t<1$, time is normalised with the sheath average radial duration 
and for $t>1$, time is normalised with the average MC radial duration.
    Orange delimits the sheath region ($0<t<1$), and blue the MC region ($1<t<4$).   
    See the caption of \fig{avrg_profiles} for more information.
    From top to bottom the parameters are the means of: 
    magnetic field $B$, bulk velocity $V$, proton density \np, proton temperature \Tp, plasma beta $\beta$ and normalised magnetic-fluctuation density $rmsBoB$, \eq{rmsBoB}.
    }
    \label{fig_split_i}
\end{figure*}

The plasma $\beta$ and the normalised magnetic fluctuations $rmsBoB$ (left/right third panels of \fig{avrg_profiles}) are the quantities that most clearly mark the MC boundaries.
Both are roughly symmetric around the MC center with a flat minimum, with $rmsBoB$ being even more symmetric than $\beta$.
$rmsBoB$ shows a maximum at the shock, and then a smooth decrease 
toward almost one-third of the MC size.

The level of absolute fluctuations $rmsB$ (right bottom panel of \fig{avrg_profiles}) shows a sharp jump with a maximum at shock arrival, while it later continuously decreases to almost one-third of the MC.
After the MC, it recovers towards typical solar wind values, but with slightly larger values. 

To summarise this section, we describe the differences between mean values inside sheaths and MCs (columns 2 and 3 of \tab{avr_values}). 
All the quantities are higher in sheaths than in MCs, by: 
$7\%$ for $V$, 8$\%$ for $B$, $\sim$ twice for $n_p$, $\sim$ 3.5 times for $T_p$, 
$\sim$ 4 times for $\beta$, $\sim$ 2.5 times for $rmsBoB$, and $\sim$ 2.6 times for $rmsB$.


\section{Sheath and MC properties for slow and fast events}
\label{sect_splits_with_V}

\subsection{Criterion to define an ordering parameter}
\label{sect_criterium_split}

The mean velocity inside the ICME at 1 AU, its mean magnetic field intensity, and its time duration could be considered as different proxies for the strength of the event.
Moreover, these values can be computed with a time window inside an ICME sub-structure such as the sheath or the MC; 
namely \Vsh, \Bsh, \dtsh, \Vmc, \Bmc, \dtmc, where the suffixes ``sh'' and ``mc'' are associated to the sheath and the MC respectively.

Which is the most relevant parameter to order the events?
To answer this question, we first preliminarily consider one of the possible strength ordering parameter, \Vmc . 
We then split our sample of 44 events into 3 subgroups with roughly the same number of events: weak (slow), mid, and strong (fast).
We produce separate superposed epoch profiles for every physical quantity analysed (similar to those shown in \fig{avrg_profiles}), 
and for each subgroup. 
Finally, we repeat this procedure for all the possible ordering parameters: 
\Bmc, \dtmc, \Vsh, \Bsh, and \dtsh.

The criterion to select the best ordering parameter is as follows.
For every ordering parameter we construct a plot as the one shown in 
\fig{Vmc_ordering} (where \Vmc\ was used in this case).
This figure shows normalised average values of all observables for each subgroup.
These values are derived from Table~\ref{tab_avr_values}.
The average value of each subgroup is shown with: red triangle up 
(events with large velocities: \Vmc $> 550$ km/s), 
green circle (mid velocities events: $450$ km/s < \Vmc $<550$ km/s), 
and blue triangle down (events with weak velocities: \Vmc $<450$ km/s).
These values are normalised with the global average using all events. 
The criterion to select which strength parameter is the best is to observe: 
i) the largest separation between mean values of the 3 subgroups, and 
ii) a coherent ordering between these values.

Among all the plots (not shown here) that we looked at, we realised that the best parameter to order the MCs is the mean MC speed \Vmc\ , which happened to be better for splitting our sample into three groups.
As is seen in \fig{Vmc_ordering}, most of the parameters are well ordered (red, green and blue in succession for each parameter).  
 The same quality of ordering and spacing as \fig{Vmc_ordering} was not found for any of the other explored ordering parameters, \eg \Bmc.  

The parameters with largest spread are the proton density \np\ , the proton temperature \Tp, the plasma $\beta$, and the GCR percentage of depression \nGCR\  (see \sect{forbush}).  
On the other hand, the parameter showing less split and no coherent order is $rmsBoB$.  
This implies that the normalised magnetic fluctuations are very similar in all groups within a $\sim 20 \%$ factor from the global mean.
Next, we observe that when the splitting is large between the groups, the relative order and spread for the sheaths are typically the same as for the MCs (\eg\ see $V$, \np, and \Tp\ in \fig{Vmc_ordering}).
Finally, the fastest MCs (red triangle up) have: the largest $B$ intensity, the lowest proton density, the hottest temperature, the lowest plasma beta, the strongest magnetic fluctuations $rmsB$, and the deepest associated GCR depression.

\subsection{Superposed epoch for each MC subset}
\label{sect_results_split}

We analyze below the averaged profiles of the three subsets.
In the upper row of \fig{split_i} we find large differences in the $B$ profiles: 
for faster/stronger events, $B$ is more peaked toward the start of the MC, while for slower/weaker events, it is more peaked toward the MC center.
Also, $B$ jumps at the shock with $\Delta B=7$, $8$, and $13$nT, for slow, mid, and fast events, respectively.

In the second row of \fig{split_i}, the ambient solar wind before the front of the sheaths is slower for weaker/slower events compared with stronger/faster events, then they are not propagating in the same mean solar wind. 
Next, the speed profiles in the three groups have all a linear profile within the MCs, which indicates a common expansion process. 
For sheaths, we also see expansion in slow events, but for fast events the sheaths have profiles showing compression.
This bimodal feature of the sheath, with an expansion and a compression, has not been reported before as far as we know, and it is shown pretty clearly in these superposed profiles with both processes having a comparable effect for MCs with mid velocities.
The fact that fast events are related to compressing sheaths is the result of the fast overtaking MC behind.
For the slow events, the magnetic field and the plasma had the time to reach a quasi-pressure balance.  
This implies an expansion similar to that of the flux rope \citep[as modeled by ][]{Demoulin09}.

In the third row of \fig{split_i}, slower events have higher proton density in all the sub-structures: sheath, MC, and its wake. 
By contrast, the ambient solar wind at the front does not show systematic changes along the three groups.  
We interpret the higher density in the sheath of slow events as a selection effect due to the drag force, as follows.
CMEs that are fast near the Sun and that later on encounter slow solar wind ahead, 
which is typically denser in the inner heliosphere \citep[\eg\ see][]{wolfe72}, accumulate more mass with a low velocity, resulting in a denser sheath and a slower MC by conservation of mechanic momentum.
Studies in this direction have been made in \cite{feng_etal15}, where they find that the SW pile-up of mass during the CME expulsion can be an important contribution to the mass increase determined by coronagraph observations.
We explore this observed feature in more details in \sect{sigma_p_sheath}.
Next, the low $n_p$ values observed in the wake of fast cases (right column of \fig{split_i}), can be due to the sweeping of ambient solar wind plasma made by faster MCs \citep[which are also typically larger, \eg , ][]{Janvier14c} combined with the difficulties to re-fill this space with new solar wind material just after the passage of the flux rope.

In the fourth row of \fig{split_i} the proton temperature in the ambient solar wind is lower for slow events, which are also traveling in a slower solar wind. 
These results are consistent because of the known direct correlation of speed and proton temperature in the quiet solar wind \citep[\eg\ ][]{Lopez86, Demoulin09d}.
Next, the faster MCs present much hotter sheaths than slower ones; this is a consequence of the local heating near the shock.
The temperature in slow MCs is lower. 
This is consistent with a relaxed flux rope which had time to adapt to the ambient solar wind pressure, to fully expand, and accordingly to cool.

\begin{figure}[ht!]
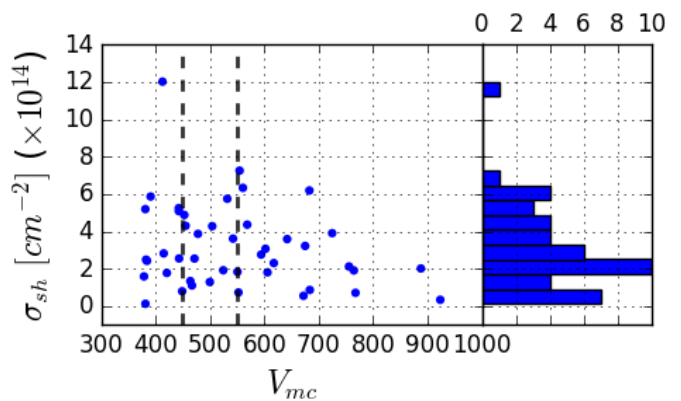

    \centering
    \img{0.73}{Vmc__vs__sigma}
    \caption{
    Left panel: Average proton surface densities of sheaths $\sigma_{sh}$, \eq{def_sigma}, as a function of the average MC speeds ($V_{mc}$) of all events. 
    The vertical dashed lines mark the thresholds between the three groups shown in \figs{split_i}{split_ii}.
    The quantity \sigsh\ is a proxy of the accumulated material ahead of MCs, within the sheaths.  
    On average, slower MCs drag more material.
    Right panel: histogram of \sigsh\ for the 44 events.
    }
\label{fig_Vmc_vs_sigmas}
\end{figure}

In the fifth row of \fig{split_i} the average plasma $\beta$ values are higher for slower events, for all ICME sub-structures.
Even in the ambient solar wind we note a tendency for slow events to travel in a less magnetised plasma (higher $\beta$). 
Within the sheaths, $\beta$ is more constant and inside the MC the $\beta$ profile is more symmetric for slow events.
For faster events $\beta$ has a more asymmetric profile, according to the asymmetry of $B$ (first row of \fig{split_i}).

In the last row of \fig{split_i} we observe big differences between the $rmsBoB$ fluctuation profiles of slow and fast events.
For instance, the fluctuation intensity is much higher near the shock for fast events.
Also, these fluctuations start even before the shock arrival and with a clearer effect for faster events.
This can be related to foreshock waves excited by energetic particles near interplanetary shocks \citep[\eg\ see,][]{blancocano_etal11}.
Next, the magnetic fluctuations have a steeper drop in the sheath-MC interface for the slow events. 
In contrast, the fluctuations are roughly the same inside the MC for the three groups.
Finally, the extension of low $rmsBoB$ values in the back suggests the presence of remnant MC-like structure due to magnetic reconnection at the front, as summarised in \sect{profiles}.

\subsection{Surface density of protons in the sheath}
\label{sect_sigma_p_sheath}

To further explore the drag as the physical reason for finding some slow events, we estimate the accumulated material inside the sheath for every event by defining the ``surface density'' \sigsh :
 \begin{equation}
 \label{eq_def_sigma}
  \sigma_{sh} = \sum_{r~\in\ {\rm sheath}} n_p \Delta r  
              = \sum_{t~\in\ {\rm sheath}} n_p V \Delta t
 \end{equation}
where $\Delta t$ is the time resolution of the data, and the sum approximates an integral along the radial direction in the ecliptic plane within the sheath boundaries.
\sigsh\ characterises the total amount of material per unit surface (perpendicular to the radial from the Sun).
\sigsh\ is associated with the material that has not escaped perpendicularly to the spacecraft crossing.
The right panel of \fig{Vmc_vs_sigmas} shows an asymmetric distribution of \sigsh, having more cases with \sigsh\ $\lesssim 3 \times 10^{14}$ cm$^{-2}$.

In the left side of \fig{Vmc_vs_sigmas} we show a scatter plot of \sigsh\ versus the average MC speeds for all the studied events.
Faster MCs tend to have less material ahead of them.
This result may seem surprising at first, since faster MCs are expected to overtake more solar wind and moreover the plasma has more difficulty to escape from the sides (faster and larger MCs).  
However, this is consistent again with the drag scenario where the MCs are slowed down when they are ejected with a high speed in the corona and encounter a slow and dense solar wind.
In contrast, high speed-MCs encountering a fast solar wind ahead are less slowed down (lower mass pile-up with faster velocity). 
These cases correspond to the lower right region of the scatter plot.
Finally, the lower left part of the plot would correspond to MCs with a slow speed encountering a low solar wind density along their travel to 1 AU.

\section{Analysis of associated Forbush events}
\label{sect_forbush}

\subsection{Superposed epoch analysis}
\label{sect_Superposed_GCR}

We analyse below the cosmic rays observed with ground-based detectors.
We use data from McMurdo neutron monitors because they 
are close enough to one geomagnetic pole, and then they can measure particles with low energies. In particular, they can observe primary protons with energies as low as $\sim$500 MeV \cite[\eg\ ][]{jordan11}.
The Larmor radius for these particles is $\sim$5x10$^{-3}$ AU, when they are embedded in an interplanetary magnetic field of $\sim$5 nT. 
For normal conditions, particles observed by these detectors are mainly of galactic origin.

An additional treatment with respect to the previous quantities 
(\sects{superposed_epoch}{splits_with_V}), is that we use the average GCR intensity before the MC sheath arrival as a reference level.
In essence, we take the data from 2 days prior the sheath arrival.
If a ground level enhancement (GLE) is present, we only take data from 6 hours prior to the sheath arrival (we verify for each event that during this time interval there is no GLE or intense noises).
Then, we normalise the GCR flux of each event with this level taken before the MC sheath defining the percentage of variation \nGCR.
Next, we continue with the same treatment as in \sect{superposed_method}.

The superposed epoch of all events is shown in the bottom left panel of \fig{avrg_profiles}.  
There is a strong decrease of \nGCR\ in the sheath followed by a progressive recovery phase during and after the MC. 
The importance of these effects depends on the MC speed as shown in the second row of \fig{split_ii}. 
The recovery time ($\tau_{FD}$) and the amplitude ($A_{FD}$) of the GCR depressions are larger for the faster events; this result is respectively consistent with the following previous studies: \cite{penna05} and \cite{richardson11}.
A minimum of \nGCR\ is present near the sheath rear for the three subsets. 
For fast events, another minimum can be observed within the fluctuations, inside the MC window and near the MC center.
This second minimum is associated to the shielding of a strong magnetic field within the flux rope configuration (with closed field lines still connected to the Sun).
This result 
is consistent with \cite{belov_etal15}, where they report that for strong-field ($>18$nT), there is a local minimum of \nGCR\ inside MCs. 

\subsection{Model for the \nGCR\ typical profiles}
\label{sect_model_nGCR}

\def \sclfg {0.20}
\begin{figure*}[ht!]
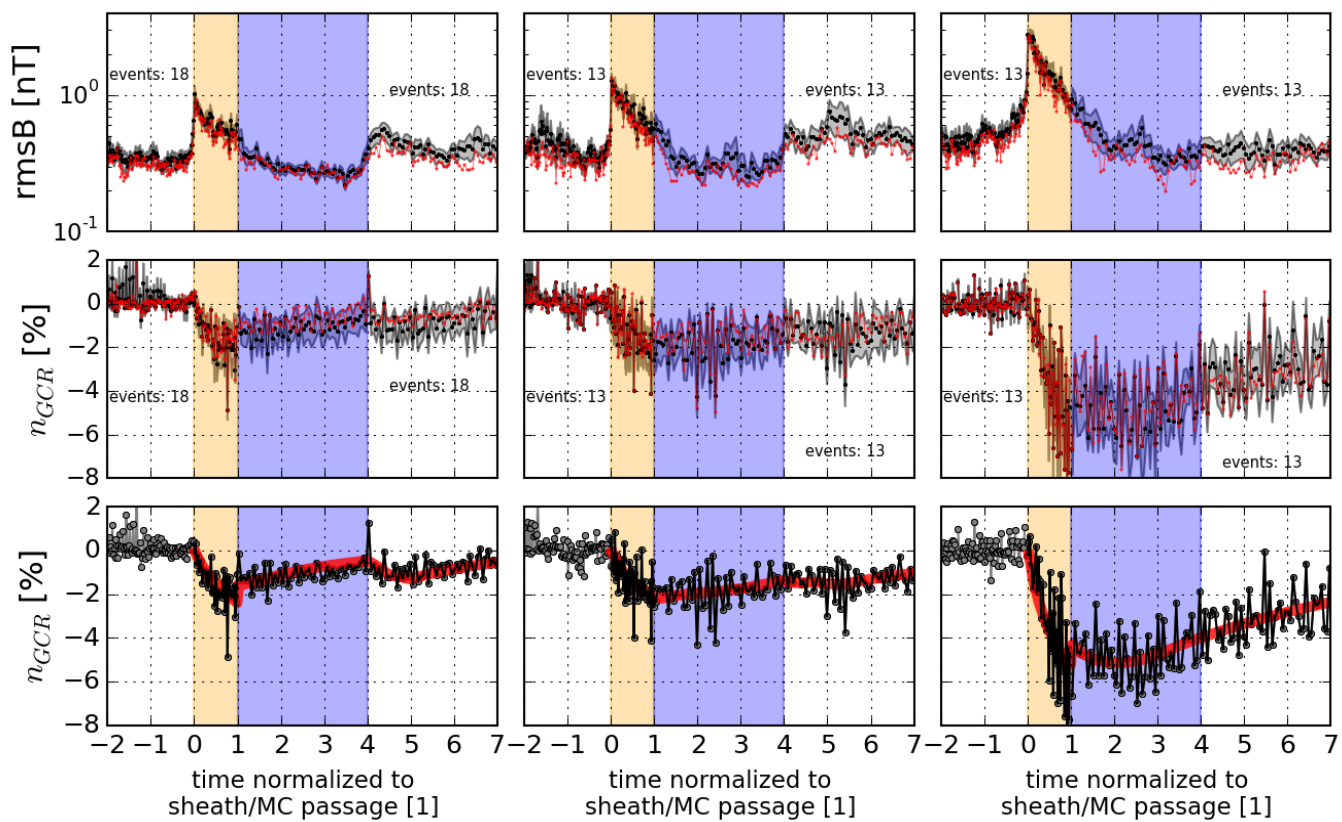

    \centering
    \img{0.8}{figs_splitted_2}
    \caption{
    Superposed epoch for the three subsets of events. Left: slow ($V<450$ km/s), middle: medium ($450$ km/s$<V<550$ km/s), and right: fast MCs ($V>550$ km/s).
    {\bf First row:} mean profiles of magnetic fluctuations $rmsB$, \eq{rmsB}.
    Typically, inside the MCs, the fluctuations are the same as in the ambient solar wind.
    For slow and mid groups, $rmsB$ has higher values in the rear with respect to the ambient solar wind.
    {\bf Second row:} mean profiles of GCR intensity \nGCR .
    Before averaging events, the intensity was normalised by the pre-shock level.  
    In the first and second rows the black line represents the average value, the grey band represents the error of the mean and the red line represents the median values in each time bin.
    {\bf Third row:} Fitted model (red lines) to the observed average profiles (see \sect{model_nGCR}).
     See the caption of \fig{avrg_profiles} for more information.
    }
    \label{fig_split_ii}
\end{figure*}

A theoretical model to describe the profile of FD was developed by \cite{wibberenz98}.
However, they did not explicitly include small-scale interplanetary magnetic fluctuations, while it was shown that these can contribute to the variety of FD profiles \citep[\eg\ ][]{jordan11}.  

In order to quantitatively analyze the processes involved in the GCR shielding produced by ICMEs, we construct a mathematical model to reproduce the GCR profiles. 
In particular, we use $B$ to account for the strong closed field lines of the MC, and $rmsB$ to account for the scattering turbulence.

From the superposed epoch profiles, we learned the following contributions (second row of \fig{split_ii}):\\
i) Inside the sheath region, \nGCR\ can be thought to be determined by an accumulation of interaction between the high energy particles and scattering centres. 
The efficiency of the scattering is related to the level of turbulent fluctuations $rmsB$.\\
ii) The decrease of \nGCR\ is larger for a stronger magnetic field inside the MC.\\
iii) The recovery mechanism is effective after the passage of the scattering turbulence region.\\
iv) There is a jump in \nGCR\ at the interface between the sheath and the MC; this can also be seen in the individual profiles. 
This feature can be related to the change of the cross diffusion coefficients associated to the plasma properties in these two different structures.

By representing each of these mechanisms with a contributing term for the temporal evolution of the GCRs flux at Earth position, we propose the following model:
\begin{align}
    \centering
    &\frac{d n_{GCR}(t)}{d t} 
     &= \text{ }& q_{\xi} \, \xi(t)  & (i)  && \nonumber \\
    &&+ \text{ }& q_{b} \, b(t)      & (ii) && \nonumber \\
    &&+ \text{ }& \frac{-1}{\tau_{FD}} n_{GCR}(t) \, \Theta(t-1) & (iii) &&\nonumber\\
    &&+ \text{ }& \Delta_o \, \delta (t-1)   & (iv) && \forall t>0
    \label{eq_FD_model}
\end{align}
with,
\begin{align}
    \xi(t) &= \textrm{max}(0, rmsB(t)-rmsB_o) \nonumber \\
    b(t)   &= \textrm{max}(0, B(t)-B_o) \,.
\end{align}
$\xi(t)$ is the $rmsB(t)$ value above its value in the quiet solar wind, $rmsB_o$, when $rmsB(t)$ is larger than $rmsB_o$. 
$b$ is the magnetic field above a reference value $B_o$ when $B>B_o$. 
$\Theta$ is the Heaviside function indicating that the recovery mechanism starts at the beginning of the MC. 
This simplification facilitates the mathematical fit and we checked that the recovery is negligible during the passage of the sheath.
$\delta$ is the Dirac delta function (more precisely distribution) which account for the observed jump of \nGCR\ at the sheath-MC interface. 
There are 5 free parameters to fit: 
$q_{\xi}$, $q_{b}$, $B_o$, $\tau_{FD}$ and $\Delta_o$.  
This model does not include explicitly the larger expansion of faster events nor a non local effect due to the sheath and MC spatial extension.  
The expansion rate is correlated to the field strength, so it is implicitly included, while we cannot include the effect of the global extension from local measurements.

In order to perform the non-linear fit of the time integral of \eq{FD_model} to the data, we employ the L-BFGS-B algorithm \citep{zhu_etal97, byrd_etal95}.
This algorithm is suitable when working with a large number of fitting variables, and it uses the gradient information of a multiple variable-function to be minimised within given bounds.
Mainly, we use this method instead of other alternatives because we observe a better convergence for our problem.

The third row of \fig{split_ii} shows the fitted models to each of the average \nGCR\ profiles, where the detailed reproduction of the observations is notable.
The inclusion of the magnetic fluctuations allows the reproduction of the observed mean decrease of \nGCR\ within the sheath with only one adjusted parameter $q_{\xi}$. 
Moreover, for the slow $V_{mc}$ group, we see a small decrease of \nGCR\ right after the MC passage, which according to our model, is due to the enhancement of $rmsB$ in the wake of the flux rope (see the left column of \fig{split_ii}). 
This behaviour is also present, but weaker, for the mid velocity group.
This result at the MC rear provides a further confirmation that magnetic fluctuations are locally affecting GCRs.
It is also in agreement with \citet[][see their Fig. 7]{badruddin_etal16}, where they find a GCR decrease around the ICME trailing edge, and simultaneously find enhanced magnetic fluctuations.
Finally, $|q_{\xi}|$ is decreasing from the low to the fast velocity group (Table~\ref{tab_fitted_vals}), which may be due to non-linearities in the diffusion process that are not taken into account in our model.

The contribution of the magnetic field $B$ is important only for the group of fast \Vmc , where \nGCR\ has a local minimum inside the MC. 
As mentioned before, this can also be related to the larger expansion of the flux rope compared to the surrounding solar wind. 
This leads to a decrease of the local density of the high energetic particles \citep[\eg\ ][]{Munakata06}.

Next, the normalised time $\tau_{FD}$, as well as the time $\tau_{FD}$ in hours (taking into account the mean MC duration of each group) are both increasing by more than a factor 2 from the slow to the fast velocity group (Table~\ref{tab_fitted_vals}).
Finally, it is worth mentioning that without the inclusion of the jump parameter $\Delta_o$, it is not possible to reproduce the recovery profile for the slow and fast $V_{mc}$ groups that clearly show this discontinuity.

%
\begin{table}
\centering
\caption{
    Fitted values of the Forbush decrease model (see \eq{FD_model}), for the three subset of events.
    These are all free parameters in the model defined by \eq{FD_model}.
    \tauFD\ is normalised to the mean duration of the MCs for each group; see \sect{model_nGCR} for informations.
    The $B_o$ value in the first column is superfluous since $q_{b}=0$, in that case. 
}
\begin{tabular}{c|ccccccccccrr}
\hline
fit parameter & \multicolumn{1}{c}{$V_{mc}^{low}$} & \multicolumn{1}{c}{$V_{mc}^{mid}$} & \multicolumn{1}{c}{$V_{mc}^{high}$} \\
\hline
$q_{\xi}$  [nT$^{-1}$]    & -9.4   & -6.0   &  -5.5    & \\
$q_{b}$  [nT$^{-1}$]      &~~0.0   & -0.9   &  -0.2    & \\
$B_o$  [nT]               &~~--    & 11.9   &  14.5    & \\
\tauFD [1]                &~~2.4   &~~4.2   &~~5.8     & \\
$\Delta_o$  [\%]          &~~0.9   &~~0.0   &~~1.0    & \\
\hline
\end{tabular}
\label{tab_fitted_vals}
\end{table}

\section{Summary and conclusions}
\label{sect_conclusions}

We obtained superposed epoch profiles during the passage of MCs and their sheaths at 1~AU, for different physical quantities observed \insitu. 
This technique allows us to identify several phenomena that are common to most of the events.
First, the magnetic field strength and the plasma velocity in the MC sheath are typically close to the values found at the beginning of the MC. 
Second, the MC has a similar density than the background solar wind, while the sheath is a factor $\approx 2.5$ denser. 

We next explored how the sheath and MC properties could be best differentiated according to the importance of the event.
We found that the mean MC velocity is the best parameter to order the sheath and MC properties. 
This provides a measure of the strength of the events.
Then, we separated our sample in three groups: slow/intermediate/fast events and compared their superposed epoch profiles.

We found that the slowest MCs have a larger proton density in their sheaths than faster events.
We next computed the density of protons per unit of surface along the sunward direction to better characterise the accumulated material that does not escape from the MC front to the sides. 
We found that on average, slow events have more massive sheaths than fast events. 
We attribute this result to a selection effect: a part of the slow MCs at 1 AU were initially fast events close to the Sun but they were slowed down as they encountered slow and dense plasma on their path to 1 AU.
Then, such decelerated MCs have more massive sheaths at 1 AU.

The comparison of the three groups show that the slow MCs have properties compatible with a more relaxed configuration.
First, they have an expanding sheath, with a very similar expansion rate as in the following MC. 
Then, the accumulated magnetic field in the sheath has the time to reach a quasi-equilibrium with the surrounding solar wind, which implies that the sheath expands with a similar rate as the driving flux rope.
Second, the slow MCs have a nearly symmetric magnetic profile around their center.
Finally, they have a lower proton temperature (as they have the time to expand enough to be in quasi-equilibrium with the surrounding solar wind).
All these characteristics together indicate that the slower MCs are expected to be in a more relaxed force-free state than faster MCs. 

The proton density and temperature are significantly larger in sheaths than in MCs. 
Still, there is no sharp transition but a progressive sheath-MC transition which extends up to 20\% of the MC duration. 
A very similar transition is also present for the amplitude of the magnetic fluctuations.  
These results are present for all three groups of events as well as in individual events, so they are not due to a smearing of sheath properties inside the MC due to inappropriate boundary definition. 
We interpret these results as the consequence of magnetic reconnection at various locations along the flux rope between sheath and flux-rope magnetic fields.
Then, while along the spacecraft trajectory the crossed magnetic field appears as belonging to the flux rope, some parts of the crossing can belong to a magnetic field region that has reconnected with the sheath magnetic field further away from the local spacecraft trajectory.
This allows plasma, heat and magnetic fluctuations to enter in this reconnected field implying, in average, sheath properties ``diffusing'' within the front of the MC.

We next studied the typical effects that these sheaths and MCs have on the cosmic ray transport by studying the associated Forbush decrease profiles.
For all MCs a local minimum of the GCR flux is present inside the sheaths, while for fast MCs another minimum is also present within the MCs. 
In the superposed epoch profiles of Forbush events, we found a discontinuity of the GCR flux at the sheath-MC interface, owing probably to the change of properties in the cross magnetic-field diffusion coefficients in these two structures.
Indeed, GCRs encounter a structure with closed and intense magnetic field lines, where the high energy particles have difficulties to penetrate. 

We finally derive a new semi-empirical model for the Earth observations of the Forbush profile that takes into account the former process 
as a simple superposition of 
(1) the enhancement of magnetic fluctuations over a threshold, 
(2) the intensity of the magnetic field over a threshold, 
(3) a jump at the time of the sheath-MC interface to consider the change of the magnetic connectivity, and 
(4) a recovery phase for times after the sheath-MC interface.
This model has five free parameters.  
We fit the model to each of the groups of events. 
Besides reproducing the first steep decrease of GCRs after the shock time and the recovery phase, we can reproduce a second decrease of GCRs in the wake of the slow MCs.
This second fall is due to the presence of scattering centres which we associate with the amplitude of the magnetic fluctuations. 
This last result is in agreement with the recent study of \cite{badruddin_etal16}.

Furthermore, our model also reproduces a local \nGCR\ minima inside the MC for the sub-group of fast events.
This feature is consistent with \cite{belov_etal15}, where they find that strong-field ($>18$ nT) are associated to local minima of \nGCR\ within MCs.

Finally, we find that the recovery time \tauFD\ is increasing by a factor a bit larger than two from the slow to the fast events.

The results presented here improve the knowledge of MCs and their sheaths, their evolution in the solar wind, and also the relationship between MCs and their GCR shielding.
In particular, the semi-empirical model for the FD profile we presented will help to improve the understanding of energetic particle transport in the heliosphere, and can be used to put constrains on theoretical models that consider 
(1) global macro-scale magnetic configurations for MCs and (2) turbulent properties of the solar wind.

\onltab{
\begin{table*}
\centering
\caption{
    List of the 44 events studied. Column: 
(1): event number,
(2): time of shock arrival \citep[as in][]{Richardson10},
(3): time difference between MC leading edge and shock time (\ie, the sheath duration),
(4): MC duration, 
(5): difference between \citeauthor{Richardson10} and the shifted shock time at L1 point, where the IP observations are compared,
(6): amplitude of the associated GCR perturbation.
}
\begin{tabular}{cccccc}
\hline
case & shock date        & sheath duration & MC duration & time shift & $A_{\rm FD}$ \\
     & [yyyy-mm-dd HH:MM]& [hours]         & [hours]     & [min]      & [\% of the background] \\
 (1) &      (2)          & (3)    & (4) & (5)  &  (6)   \\
\hline
   1 & 1998-03-04 11:56  &  ~~1.1 &  41 &  -60 & ~~-0.6 \\
   2 & 1998-05-01 21:56  &   14.1 &  29 &  -30 & ~~-6.9 \\
   3 & 1998-09-24 23:45  &   10.2 &  27 &  -30 &  -13.3 \\
   4 & 1998-10-18 19:52  &  ~~8.1 &  10 &  -50 & ~~-2.1 \\
   5 & 1998-11-08 04:51  &   20.1 &  24 &  -30 & ~~-6.5 \\
   6 & 1999-02-18 02:46  &   11.2 &  22 &  -40 & ~~-5.6 \\
   7 & 1999-04-16 11:25  &  ~~8.6 &  25 &  -45 & ~~-0.6 \\
   8 & 1999-08-08 18:41  &   26.3 &  20 &  -60 & ~~-0.1 \\
   9 & 2000-02-11 23:52  &   17.1 &   7 &  -40 & ~~-4.4 \\
  10 & 2000-02-20 21:39  &   12.3 &  26 &  -60 & ~~-1.6 \\
  11 & 2000-06-23 13:03  &   13.9 &  41 &  -40 & ~~-2.3 \\
  12 & 2000-07-28 06:34  &   14.4 &  13 &  -50 & ~~-1.6 \\
  13 & 2000-08-11 18:45  &   10.2 &  24 &  -40 & ~~-2.9 \\
  14 & 2000-09-17 17:57  &  ~~8.1 &  16 &  -60 & ~~-5.2 \\
  15 & 2000-10-03 00:54  &   16.1 &  21 &  -45 & ~~-1.9 \\
  16 & 2000-10-28 09:54  &   11.1 &  25 &  -50 & ~~-6.5 \\
  17 & 2000-11-06 09:48  &   12.2 &  20 &  -35 & ~~-6.4 \\
  18 & 2001-04-04 14:55  &  ~~3.1 &  14 &  -40 & ~~-4.6 \\
  19 & 2001-04-11 13:43  &   18.3 &  10 &  -30 & ~~-9.9 \\
  20 & 2001-04-21 16:01  &  ~~7.0 &  26 &  -30 & ~~-1.3 \\
  21 & 2001-04-28 05:01  &   21.0 &  11 &  -30 & ~~-7.7 \\
  22 & 2001-05-27 14:59  &   21.0 &  22 &  -45 & ~~-3.5 \\
  23 & 2001-10-31 13:48  &  ~~6.2 &  38 &  -60 &~~~~0.4 \\
  24 & 2002-03-18 13:22  &   33.6 &  17 &  -60 & ~~-4.8 \\
  25 & 2002-03-23 11:37  &   24.4 &  34 &  -45 & ~~-2.9 \\
  26 & 2002-04-17 11:07  &   15.9 &  23 &  -45 & ~~-4.9 \\
  27 & 2002-04-19 08:35  &   27.4 &  30 &  -30 & ~~-3.3 \\
  28 & 2002-05-23 10:50  &   12.2 &  18 &  -40 & ~~-5.8 \\
  29 & 2002-08-01 05:10  &  ~~6.8 &  11 &  -50 & ~~-1.3 \\
  30 & 2002-08-01 23:09  &  ~~9.8 &  12 &  -50 & ~~-3.4 \\
  31 & 2003-03-20 04:40  &  ~~7.3 &  10 &  -25 & ~~-2.5 \\
  32 & 2003-08-17 14:21  &   20.6 &  18 &  -40 & ~~-1.9 \\
  33 & 2003-11-20 08:03  &  ~~1.9 &  16 &  -35 & ~~-5.0 \\
  34 & 2004-04-03 10:00  &   16.0 &  37 &  -60 & ~~-2.0 \\
  35 & 2004-07-22 10:36  &  ~~4.4 &   6 &  -45 & ~~-4.6 \\
  36 & 2004-07-26 22:49  &  ~~3.2 &  10 &  -25 &  -13.3 \\
  37 & 2004-08-29 09:09  &  ~~9.8 &  27 &  -50 & ~~-0.6 \\
  38 & 2004-11-07 18:27  &  ~~7.5 &  15 &  -30 & ~~-7.9 \\
  39 & 2005-05-20 03:00  &  ~~4.0 &  22 &  -60 &~~~~0.3 \\
  40 & 2005-06-12 07:45  &  ~~7.2 &  16 &  -60 & ~~-3.0 \\
  41 & 2005-06-14 18:35  &   10.4 &  28 &  -45 & ~~-0.8 \\
  42 & 2005-07-17 01:34  &   12.4 &  14 &  -60 & ~~-4.5 \\
  43 & 2005-12-31 00:00  &   13.0 &  22 &  -40 & ~~-2.1 \\
  44 & 2006-12-14 14:14  &  ~~7.8 &  22 &  -25 & ~~-8.0 \\
\hline
\end{tabular}
\label{tab_selec_events}
\end{table*}
}

\begin{acknowledgements}
Thanks to the public service of NASA Space Physics Data Facility (SPDF) for the observations.
We thank Roger Clay and the anonymous referee for reading carefully the manuscript, and their valuable comments. 
This work was partially supported by the Argentine grants UBACyT 20020120100220 (UBA), PICT-2013-1462 (FONCyT-ANPCyT), 
PIP-11220130100439CO (CONICET) and PIDDEF 2014-2017 nro. 8 (Ministerio de Defensa, Argentina), 
and by a one month invitation of SD by Paris Observatory.
JJMM, SD, and LR acknowledge the cooperation project BE/13/07 (F.R.S.-FNRS/MinCyT) between Belgium and Argentina.
S.D. is member of the Carrera del Investigador Cien\-t\'\i fi\-co, CONICET. J.J. Mas\'\i as-Meza is a fellow of CONICET.
L. R. acknowledges support from the Belgian Federal Science Policy Office through the ESA-PRODEX program. 
This research has been partially funded by the Interuniversity Attraction Poles Programme initiated by the Belgian Science Policy Office (IAP P7/08 CHARM). 
\end{acknowledgements}

%
   \bibliographystyle{aa} 
   \bibliography{mc} 
%

\end{document}